\documentclass[conference]{IEEEtran}
\IEEEoverridecommandlockouts
\usepackage{cite}
\usepackage{amsmath,amssymb,amsfonts}
\usepackage{algorithmic}
\usepackage{graphicx}
\usepackage{textcomp}
\usepackage{xcolor}
\usepackage{multirow}
\usepackage{float}
\def\BibTeX{{\rm B\kern-.05em{\sc i\kern-.025em b}\kern-.08em
    T\kern-.1667em\lower.7ex\hbox{E}\kern-.125emX}}
\begin{document}

\title{FastGraphTTS: An Ultrafast Syntax-Aware Speech Synthesis Framework\\
}

\author{\IEEEauthorblockN{Jianzong Wang, Xulong Zhang$^{\ast}$\thanks{$^{\ast}$Corresponding author: Xulong Zhang (zhangxulong@ieee.org)}, Aolan Sun, Ning Cheng, Jing Xiao}
\IEEEauthorblockA{\textit{Ping An Technology (Shenzhen) Co., Ltd., Shenzhen, China} }
}

% \author{\IEEEauthorblockN{Kaiyi Luo$^{1,2\dagger}$, Xulong Zhang$^{1\dagger}$\thanks{$^\dagger$ Both authors have made equal contributions.}, Jianzong Wang$^{1\ast}$\thanks{$^{\ast}$Corresponding author: Jianzong Wang (jzwang@188.com)}, Huaxiong Li$^{2}$, Ning Cheng$^{1}$, Jing Xiao$^{1}$}
% \IEEEauthorblockA{\textit{$^{1}$Ping An Technology (Shenzhen) Co., Ltd. } \\ 
% \textit{$^{2}$University}
% }
% }

% \author{\IEEEauthorblockN{Yazhong Si$^{1,2\dagger}$, Xulong Zhang$^{1\dagger}$\thanks{$^\dagger$ Both authors have made equal contributions.}, Fan Yang$^{2}$, Jianzong Wang$^{1\ast}$\thanks{$^{\ast}$Corresponding author: Jianzong Wang (jzwang@188.com)},  Ning Cheng$^{1}$, Jing Xiao$^{1}$}
% \IEEEauthorblockA{\textit{$^{1}$Ping An Technology (Shenzhen) Co., Ltd. } \\ 
% \textit{$^{2}$University}
% }
% }

\maketitle

\begin{abstract}
    This paper integrates graph-to-sequence into an end-to-end text-to-speech framework for syntax-aware modelling with syntactic information of input text. Specifically, the input text is parsed by a dependency parsing module to form a syntactic graph. The syntactic graph is then encoded by a graph encoder to extract the syntactic hidden information, which is concatenated with phoneme embedding and input to the alignment and flow-based decoding modules to generate the raw audio waveform. The model is experimented on two languages, English and Mandarin, using single-speaker, few samples of target speakers, and multi-speaker datasets, respectively. Experimental results show better prosodic consistency performance between input text and generated audio, and also get higher scores in the subjective prosodic evaluation, and show the ability of voice conversion. Besides, the efficiency of the model is largely boosted through the design of the AI chip operator with 5x acceleration.
\end{abstract}


\section{Introduction}
\label{sec:intro}
% 问题，进展
Text-to-speech system is an essential module of human-computer interaction system. The system is designed to consume text input and output synthesised audio. Considering the high-resolution feature of speech audio, majority of academics convert the samples from the temporal domain into the spectral domain to reduce the modelling dimension and extract salient features \cite{ cho22_interspeech, ju22_interspeech, udagawa22_interspeech, ammarabbas22_interspeech}. Linear or mel-spectrograms are proposed to represent speech features in more compact spaces \cite{yang2021multi}. Mel spectrograms are processed by applying a mel basis on linear spectrograms to highlight features in the frequency range where humans are more sensitive. However, such spectral features lose phase information of speech, which may lead to loss of information in speech reconstruction. Furthermore, the strong assumption that mel spectrograms can represent the speech to be modelled compromises the ability of deep learning models to model high-dimensional features. So the ultimate goal of text-to-speech is to find a paradigm that can map from text to raw audio. In this case, a high-resolution audio decoder is required, and a text encoder needs to be designed. 

% 现阶段工作
There have been few attempts to generate raw audio from raw text input. EATS \cite{donahue2021endtoend} uses Generative Adversarial Networks to generate raw audio through adversarial learning. VITS \cite{DBLP:journals/corr/abs-2106-06103} combines the concepts of glow and variational autoencoder (VAE), and proposes a monotonic alignment search (MAS) algorithm, which makes end-to-end learning non-autoregressive. Wave-Tacotron \cite{DBLP:journals/corr/abs-2011-03568} makes minor modifications to Tacotron to generate raw audio directly by incorporating normalizing flows into an autoregressive decoder loop. Some other scholars have tried to apply diffusion methods to text-to-speech, such as WaveGrad \cite{chen2021wavegrad}, WaveGrad-2 \cite{50409}, possesses entirely differentiable and streamlined designs capable of directly producing audio waveforms without the need for generating spectrograms or any other intermediary elements. Furthermore, non-autoregressive networks like EfficientTTS \cite{pmlr-v139-miao21a} and Fastspeech 2s \cite{DBLP:conf/iclr/0006H0QZZL21} become feasible with the addition of audio vocoders at the end for end-to-end learning. In fact, most GAN-based models successfully model raw audio with multi-resolution discriminators, while other non-autoregressive models add a vocoder at the end for joint training. Furthermore, reinforcement learning methods have been applied to model the most difficult text-to-speech modules, aligning text and speech to facilitate end-to-end speech synthesis \cite{Reinforce-Aligner}. 

% 这些工作存在的问题
However, when re-experimenting with the above methods, a common problem is that syntactic information is not utilized in speech synthesis, resulting in prosody inconsistency between the input text and output speech, that is, the existing models have certain randomness in stress and pauses of synthesized audio, regardless of the syntactic structure of the input text. Some researchers have attempted to use the syntactic structure of the textual input to relate the prosody of the output audio to the syntactic information of textual input \cite{10022793}. \cite{DBLP:journals/corr/abs-1904-04764} first attempts to exploit syntactic features by dependency parsing trees to improve acoustic models. GraphTTS \cite{DBLP:conf/icassp/SunWCPZX20} uses graph neural networks to parse graph embeddings to model syntactic relations. GraphSpeech \cite{DBLP:conf/icassp/0008S021} made another attempt to exploit a similar structure on Transformer networks. GraphPB \cite{DBLP:conf/slt/SunWCPZKX21} attempts to graphically model prosodic boundaries and analyse them by graph neural networks. Afterwards, \cite{DBLP:conf/icassp/SongLZ0M21} conducted more experiments on the utilization and modeling of syntactic graphs. Incorporating tree-structured syntactic data into the prosody modeling component of PortaSpeech~\cite{ren2021portaspeech} is a key feature of SyntaSpeech~\cite{ye2022syntaspeech}. Inspired by DialogueGCN \cite{ghosal-etal-2019-dialoguegcn}, \cite{9747837} introduces a multi-modal context modeling approach based on graphs and applies it to conversational text-to-speech (TTS) systems to improve the expressive qualities of the synthesized speech. However, none of the above methods attempt to solve the true end-to-end problem, as they all require a vocoder to restore the original audio information.

% 我们的工作
Therefore, in this paper, syntactic information is utilized for text modeling and input to a graph neural network followed by a speech decoder to achieve end-to-end. GNNs can capture the syntactic information of the input text and improve the prosody consistency of the synthesized speech, but it also require high computational complexity and memory bandwidth. Therefore, a novel AI chip operator was designed that leverages the parallelism and flexibility of the AI chip architecture, which features a large on-chip SRAM, multiple independent processor cores, and a bulk synchronous parallel execution model. Contributions are as follows:
\begin{itemize}
      \item Syntactic information is used in an end-to-end text-to-speech framework to improve prosody consistency between input text and generated speech, resulting in a more reasonable audio prosodic rhythm;
      \item Few-shot samples of target speakers are used to clone the timbre through the transfer learning technique;
      \item Multi-speaker text-to-speech and many-to-many voice conversion are implemented to improve voice diversity;
      \item For efficient large-scale parallel computing, specific AI chip operators are proposed to further improve computing efficiency.
      % we utilize computing architecture of near memory (900M SRAM on chip) and mass cores(1472 cores), distributed on-chip memory (53TB/s memory access bandwidth) and efficient inter-core communication (11TB/s exchange bandwidth) can further improve computing efficiency.
\end{itemize}

\section{Related Works}
\label{sec:related}

\subsection{Neural TTS}

Natural Text-to-Speech (TTS)~\cite{wang2017tacotron,DBLP:conf/nips/RenRTQZZL19} has always been one of the highly regarded research directions in the field of artificial intelligence. With the continuous development of deep learning technology, Neural TTS has emerged as a significant breakthrough in the TTS domain. Neural TTS is a neural network-based method for speech synthesis, and its basic principle involves training deep neural networks to convert text information into natural and fluent speech.

Zhang et al.~\cite{zhang2022semi} introduced a semi-supervised learning approach for neural text-to-speech (TTS) in resource-constrained settings, addressing limited labeled target data. It leverages a pre-trained reference model using Fastspeech2~\cite{DBLP:conf/iclr/0006H0QZZL21} to produce pseudo labels for the target data, subsequently refining the model through a dual loss framework comprising hard loss and reference loss. 

Zhang et al.~\cite{zhang2022adapitch} introduced Adapitch, a multi-speaker text-to-speech (TTS) approach that adapts to untranscribed data by employing self-supervised text-to-text and mel-to-mel modules. It also incorporates a supervised content disentangling module to separate pitch, text content, and speaker identity. 
To address challenges in emotional speech synthesis, Tang et al.~\cite{tang2023qi} proposed QI-TTS model, which extends beyond existing models to capture fine-grained prosody control, including intonation variations. 

Wang et al.~\cite{wang2023sar} introduced a self-supervised learning framework utilizing autoencoders, incorporating a distortion prior to acquire acoustic representations (SAR) resistant to distortion, replacing manually designed Mel-spectrograms. This distortion prior involves randomly masking parts of the autoencoder's latent space features at varying proportions, compelling it to grasp high-level phonetic structures and enabling the inference of missing data from the remaining features. These learned acoustic representations are subsequently employed to train neural vocoders and acoustic models, forming an end-to-end text-to-speech system. 

Tang et al. proposed EmoMix~\cite{tang2023EmoMix}, which is an emotion-aware speech synthesis framework based on diffusion probability models and pretrained emotion recognition models. It can generate speech with specified intensity or blended emotions. Emotion embeddings are extracted using a pretrained emotion recognition model from reference speech, serving as additional conditions for the diffusion probability model to generate primary emotion categories. 

Compared to traditional TTS methods, Neural TTS offers several significant advantages. The speech generated by Neural TTS is usually more natural and fluent, closely resembling human pronunciation. Different styles and tones of TTS models can be trained to meet various scenarios and requirements. Neural TTS is also easier to build as an end-to-end model, simplifying the system architecture.

\subsection{Graph Neural Network}

Graph Neural Networks (GNNs) are a category of deep learning models used for processing graph data, and they have made significant progress in recent years~\cite{DBLP:conf/aaai/YuHSX22,DBLP:conf/nips/AamandCINRSSW22}. They excel not only in domains such as social networks, bioinformatics, and recommendation systems but also have had a significant impact on traditional deep learning tasks like computer vision and natural language processing~\cite{velivckovic2023everything}. Graphs are complex data structures composed of nodes and edges, used to represent relationships between entities. Graph data is commonly found in social networks, transportation networks, protein-protein interaction networks, and more. Unlike traditional matrix data, the topological structure of graph data~\cite{wu2022graph} is crucial for analysis and prediction.

The Graph Convolutional Layer~\cite{cao2022applications} is one of the core components of GNNs, allowing nodes to update their representations by leveraging information from neighboring nodes. The design of the Graph Convolutional Layer has become a cornerstone for subsequent research. GNNs aim to map each node into a lower-dimensional vector space for subsequent tasks such as node classification and link prediction. To capture differences in importance between different nodes, researchers introduced graph attention mechanisms, enabling GNNs to assign different weights to neighboring nodes during the information aggregation process.

The earliest GNNs were based on spectral methods~\cite{he2022msgnn} using the graph Laplacian matrix, but they had high computational complexity. Later, graph convolutional layers based on neighboring node information were introduced, significantly improving efficiency. The introduction of attention mechanisms, inspired by Bahdanau et al.~\cite{BahdanauCB14}, led to the incorporation of graph attention mechanisms, allowing GNNs to better handle large-scale graph data. Researchers have proposed many variations and extensions, including multi-layer GNNs, Graph Convolutional Networks (GCN), to adapt to different tasks and application scenarios.

GraphTTS~\cite{DBLP:conf/icassp/SunWCPZX20} is the fisrt application of GNN to TTS task. It is a graph-to-sequence model for neural text-to-speech (TTS), featuring a Graph Auxiliary Encoder (GAE) module for prosody enhancement. It utilizes character-level graph embeddings to capture phonetic and syntactic information, outperforming existing sequence-to-sequence models in naturalness and prosody quality across English and Chinese datasets, all without manual audio reference selection.

Sun et al.~\cite{DBLP:conf/slt/SunWCPZKX21} proposed Graphical Prosody Boundary (GraphPB) approach in Chinese speech synthesis, utilizing graphical representations to enhance prosody performance by capturing semantic and syntactic relationships. It introduces two methods for constructing graph embeddings based on hierarchical prosody boundaries and integrates sequential information into a graph-to-sequence text-to-speech model.

\subsection{AI Chip Operator}

AI chip operator aims to design and optimize specialized hardware operators for AI applications, such as GNNs and TTS. AI chip operator leverages the parallelism and flexibility of the AI chip architecture, which consists of multiple processing elements (PEs) connected by a reconfigurable network-on-chip (NoC). AI chip operator can dynamically map the computation graph of an AI model onto the AI chip, by assigning different PEs to different nodes and edges of the graph, and configuring the NoC to route the data between PEs. AI chip operator can also adapt to different AI models and workloads by changing the mapping scheme and the NoC configuration.

Wu et al.~\cite{wu2023turbomgnn} proposed a novel fine-grained kernel fusion method for improving the efficiency of concurrent GNN training tasks on GPU. A concurrent GNN training framework called TurboMGNN was implemented based on the proposed method and evaluation results show that TurboMGNN can achieve up to 2.6x speedup over the state-of-the-art GNN training systems.

SWITCHBLADE~\cite{lu2023accelerating} is a framework for accelerating graph neural networks (GNNs) that addresses the challenges of high bandwidth demand and high model variety. SWITCHBLADE proposes three generic methods that span algorithmic, software, and hardware aspects: partition-level operator fusion (PLOF), shard-level multi-threading (SLMT), and fine-grained graph partitioning (FGGP). These methods reduce off-chip memory access, enhance hardware utilization, and improve data locality for various GNN models. SWITCHBLADE consists of a compiler, a graph partitioner, and a hardware accelerator that implement the proposed methods. The paper evaluates SWITCHBLADE on different GNN models and datasets and shows that it achieves significant speedup and energy savings compared to GPUs and comparable performance to model-specific GNN accelerators.

\section{Methodology}
\label{sec:method}
This section presents the details of the proposed method, which is divided into three main modules, the end-to-end FastGraphTTS framework, the details of the graph encoder, and the details of the AI chip operator architecture.

\begin{figure}[t]

\begin{minipage}[t]{1.0\linewidth}
  \centering
  \centerline{\includegraphics[width=0.75\linewidth]{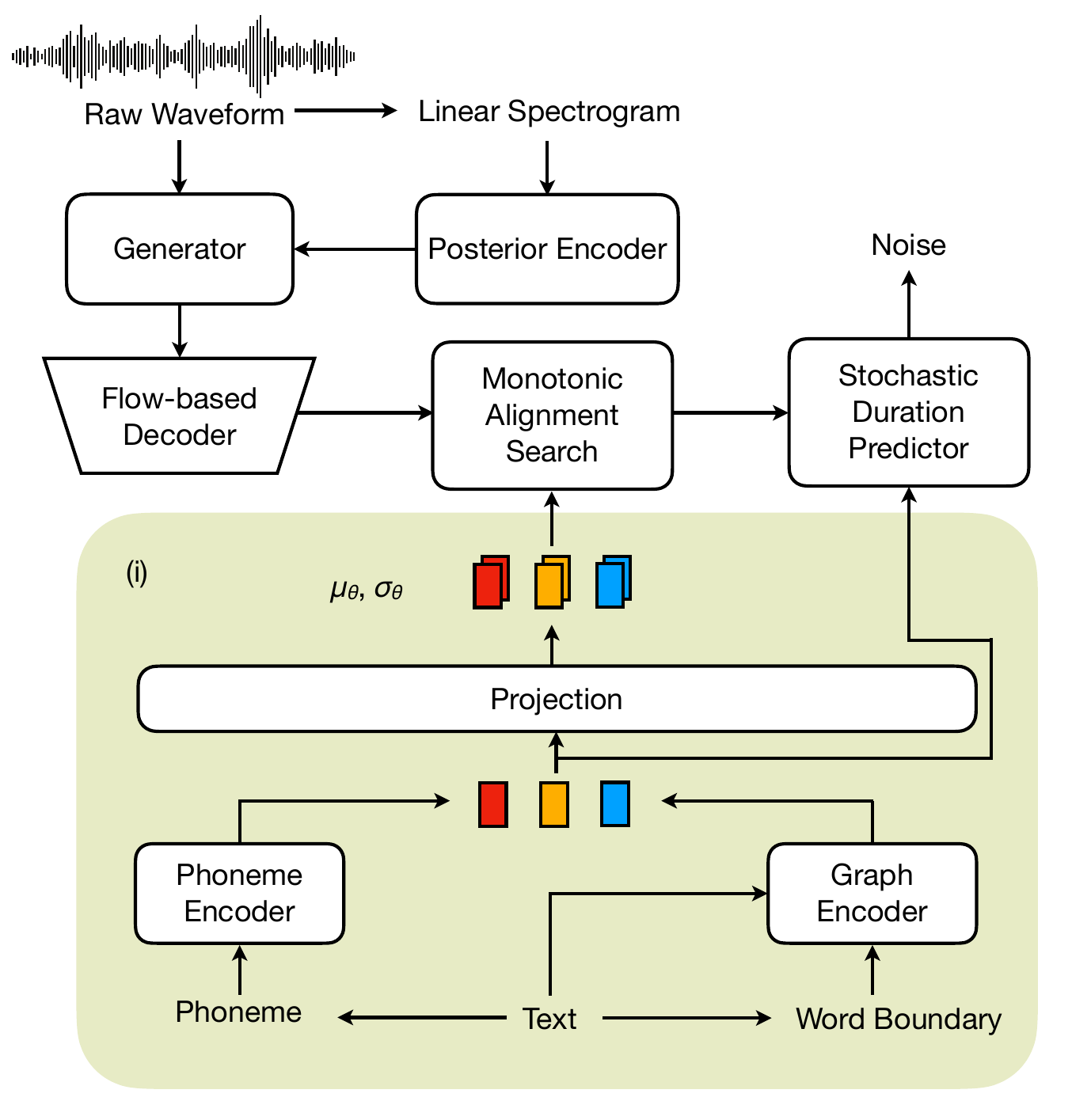}}
  \centerline{(a) The framework of FastGraphTTS} % \medskip
\end{minipage}
\\
\\
\begin{minipage}[t]{1.0\linewidth}
  \centering
  \centerline{\includegraphics[width=0.75\linewidth]{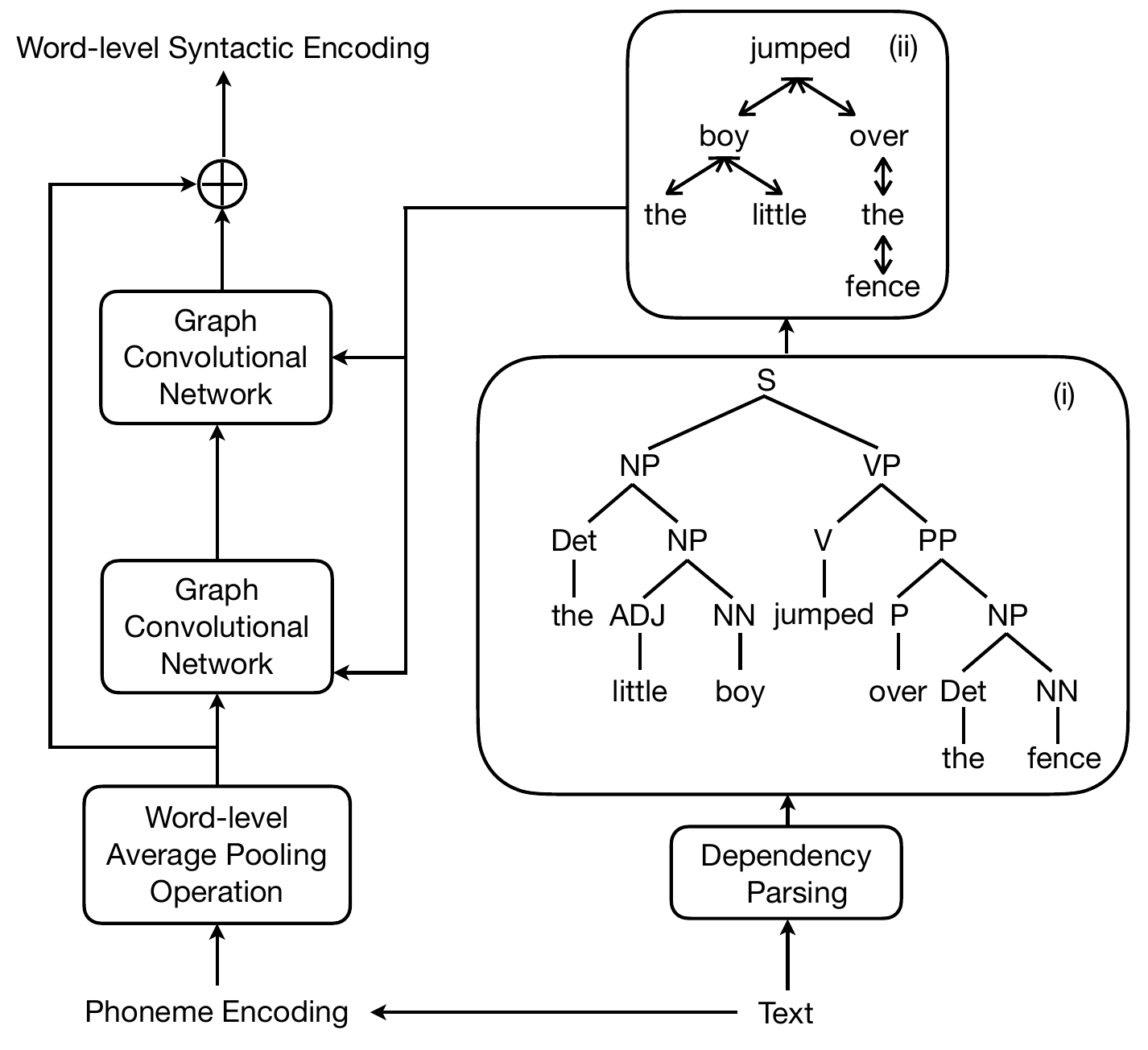}}
  \centerline{(b) Details of graph encoder.} % \medskip
\end{minipage}

\caption{End-to-end FastGraphTTS framework}
\label{fig:e2egraphtts}
\end{figure}

\subsection{End-to-End FastGraphTTS Framework}
\label{subsec:method1}
The end-to-end FastGraphTTS framework is shown in Figure \ref{fig:e2egraphtts}. The input to the framework can be raw text or even graphic languages such as Chinese and Japanese, and the model produces a sound wave as its output. During training, the text is input to the graph encoder to output the graph hidden state vector $g_{text}$. At the same time, the input text is converted into phonemes, which are consumed by the phoneme encoder to output the phoneme embedding $p_{text}$. $g_{text}$ and $p_{text}$ are concatenated and projected into the mean and variance $\mu_{\theta}$ and $\sigma_{\theta}$ for the following normalising flow process. Meanwhile, the linear spectrogram is extracted from the original audio and encoded by the posterior encoder, which outputs the spectrogram hidden state $z$. $z$ is input to the flow $f_{\theta}$ and produces $f_{\theta}(z)$. A monotonic alignment search (MAS) is performed between $f_{\theta}(z)$ and $\mu_{\theta}$ and $\sigma_{\theta}$. MAS then generates a duration value $d$ which is input to the stochastic duration predictor and generates noise. At the same time, $z$ is divided into several pieces, which are input to the generator to output the waveform.

In the inference process, the text to be inferred is input to the graph encoder, encoded as $g_{text}$, and the phoneme embedding $p_{text}$ is generated at the same time, and the concatenated vector of the two is projected to $\mu_{\theta}$ and $\sigma_{\theta}$. Meanwhile, the noise is predicted by a stochastic duration predictor to predict the duration, and it is mapped to $\mu_{\theta}$ and $\sigma_{\theta}$ to generate $f_{\theta}(z)$. This is converted by the inverse flow $f_{\theta}^{-1}$ and outputs $z$, decoded as a waveform.

\subsection{Details of Graph Encoder}
\label{method2}
\subsubsection{Graph embedding}
Different forms of graph embeddings have been proposed in previous studies, such as phoneme-level graph embeddings where edges are represented by sequence information, word-level graph embeddings where edges are represented by syntactic parsed trees, and graph embeddings where edges are represented by prosodic boundaries \cite{DBLP:conf/icassp/SunWCPZX20, DBLP:conf/slt/SunWCPZKX21}. In this model, word-level graph embeddings are obtained through dependency parsing analysis as model input.

\begin{table*}[t]
    % 1
\caption{Details of the datasets used in the experiments}
\begin{tabular}{ccccccccc}
\hline
\textbf{Dataset} & \textbf{Language}&\textbf{Sample rate}  & \textbf{\#speakers} & \textbf{\#hours} & \textbf{\#utterances}  & \textbf{\#train} & \textbf{\#validation} & \textbf{\#test} \\ \hline
LJSpeech & English & 22KHz  & 1& 24 & 13100  & 12500 & 100 & 500\\
BIAOBEI & Mandarin& 48KHz  & 1 & 12  & 10000  & 8000 & 1000 & 1000\\
Target speaker A  & English & 22KHz & 1 & 0.1& 30  & 30 & 10 & 500\\
Target speaker B  & Mandarin & 22KHz& 1 & 0.1& 30   & 30 & 10 & 500\\
VCTK  & English & 44KHz & 109 & 44 & 44000  & 35200 & 4400 & 4400 \\
AISHELL-3  & Mandarin & 44KHz & 218 & 85 & 88035 & 70428 & 8803 & 8804\\\hline
\end{tabular}
\centering
\label{tab:details of dataset}
\end{table*}

\subsubsection{Architecture of graph encoder}
The graph encoder’s intricate design is displayed in Fig.\ref{fig:e2egraphtts} (b). The input text is first converted to phoneme encoding according to the text processors of different languages. Phoneme encoding is introduced into the word-level average pooling (WP) operation for information averaging. The output of the WP becomes the input for a two-layer graph convolutional network (GCN) afterwards for message passing and aggregation. At the same time, the input text is parsed into a syntax tree as shown in Figure \ref{fig:e2egraphtts} (b) (i) through a dependency parse tree. The forward and backward edges of the syntax tree are used as bi-directional graph edges to generate the syntax graph shown in Figure \ref{fig:e2egraphtts} (b) (ii). The syntactic graph is input to two layers of graph convolutional network (GCN). The result of WP and the result of GCN are combined together, and the word-level syntactic embedding is output, which is then input to the projection layer.

\subsubsection{Graph convolutional network}
This paper adopts the most typical graph neural network, the graph convolutional network (GCN). GCN is a convolutional neural network (CNN)-like architecture that is suitable for non-Euclidean data, such as molecular data or social network data. It can also handle syntactic structures we build from time series data.

% \begin{equation}
%     \mathcal{L} = \mathcal{L}_0 + \lambda\mathcal{L}_{reg}, 
% \label{eq1}
% \end{equation}

% \begin{equation}
%     \mathcal{L}_{reg} = \sum_{i,j}A_{ij}||f(X_i) - f(X_j) ||^2 = f(X)^T \Delta f(X)
% \label{eq2}
% \end{equation}

\begin{equation}
    \begin{aligned}
    \mathcal{L} &= \mathcal{L}_0 + \lambda\sum_{i,j}A_{ij}||f(X_i) - f(X_j) ||^2\\ &= \mathcal{L}_0 +f(X)^T \Delta f(X)\\
\label{eq1-2}
\end{aligned}
\end{equation}
here, in the equation \ref{eq1-2}, $\mathcal{L}_0$ represents the supervised loss of the model and $\lambda$ is the weighting factor. In the equation \ref{eq1-2}, $f(\cdot)$ can be a differentiable function like a neural network, $X$ represents a matrix comprising feature vectors denoted as $X_i$ associated with the individual nodes within the graph.

The symbol $\Delta$, denoting the unnormalized graph Laplacian, emerges from the interplay of elements within an undirected graph $\mathcal{G} = (\mathcal{V}, \mathcal{E})$, characterized by a set of $N$ nodes represented as $v_i \in \mathcal{V}$ and edges denoted as $(v_i, v_j) \in \mathcal{E}$. Within this framework, $A$ is an $N \times N$ adjacency matrix capturing the pairwise connections between nodes, and $D_{ii}$ is defined as the sum of elements in the $i$-th row of $A$, effectively forming the degree matrix. Multilayer Graph Convolutional Networks (GCNs) follow layer-wise propagation rule as follows:
\begin{equation}
    H^{(l+1)} = \sigma (\Tilde{D}^{-\frac{1}{2}}\Tilde{A}\Tilde{D}^{-\frac{1}{2}}H^{(l)}W^{(l)})
\end{equation}
the modified adjacency matrix $\Tilde{A}$ is derived by augmenting the original adjacency matrix $A$ of the undirected graph $\mathcal{G}$ with self-connections, achieved through the addition of the identity matrix $I_N$. Notably, $I_N$ stands as the identity matrix of size $N \times N$. Moreover, $\Tilde{D}_{ii}$, calculated as the summation of elements in the $i$-th row of $\Tilde{A}$, constitutes the degree matrix with self-connections included.

Within the neural network framework, $W^{(l)}$ represents a layer-specific trainable weight matrix, while $\sigma(\cdot)$ denotes an activation function, commonly exemplified by $ReLU(\cdot) = \max(0, \cdot)$. The matrix $H^{(l)}$, where $l$ signifies the layer index, resides in $\mathbb{R}^{N \times D}$ and encapsulates the activations within the $l^{th}$ layer, with the initial activation matrix $H^{(0)}$ being equivalent to the input matrix $X$.

The forward model is elegantly expressed in a straightforward manner as follows:
\begin{equation}
Z = f(X, A) = softmax(\hat A ReLU (\hat AXW^{(0)}) W^{(1)})
\end{equation}

We introduce two essential weight matrices for our neural network architecture:
\begin{itemize}
\item $W^{(0)} \in \mathbb{R}^{C \times H}$, serving as the input-to-hidden weight matrix, specifically designed for a hidden layer comprising $H$ feature maps.

\item $W^{(1)} \in \mathbb{R}^{C \times H}$, which functions as the hidden-to-output weight matrix, facilitating the transformation of hidden layer activations into the final output.
\end{itemize}

To perform the activation, we employ the softmax function, mathematically defined as $softmax(x_i) = \frac{1}{z} \exp(x_i)$, where $z$ represents the normalization constant derived from the summation of exponentiated values within the row, ensuring the row-wise application of the softmax activation.

The propagation of GCN can be summarised as follows:
\begin{equation}
    h_v^{(l)} = \sigma (\sum_{u \in N(v)} W^{l} \frac{h_u^{l-1}}{|N(v)|} )
\end{equation}

\subsection{Details of AI Chip Operator}

The design of this AI chip is characterized by its remarkable degree of parallelism, boasting an impressive 1472 individual compute cores referred to as `tiles.' Each of these tiles is equipped with its own local memory capacity, totaling 642KB, and is further endowed with the capability to concurrently run 6 hardware threads, each capable of executing distinct programs independently. In stark contrast, conventional GPUs feature a rather limited cache directly on the chip, necessitating the retrieval of all data from off-chip DRAM or High Bandwidth Memory (HBM), and providing comparatively less flexibility when it comes to executing diverse programs across individual threads.

To harness the inherent parallelism of the AI chip effectively, it leverages the BSP execution model, known as Bulk Synchronous Parallelism. In this model, all the tiles operate in parallel, leveraging their individual local memory resources. As each tile completes its computation tasks, it transitions into a brief waiting phase, effectively going idle. Only when all tiles have completed their respective computations, a concise synchronization phase occurs, involving all the tiles, ensuring that they are collectively prepared for the next step. Subsequently, data transfer between tiles is executed with remarkably high bandwidth, adhering to a predefined schedule, during the exchange phase. This cyclical process then recommences, as all tiles re-enter the compute phase. It's noteworthy that the duration between synchronization phases isn't fixed but rather dynamically determined by the time required for the ongoing computation, allowing for adaptability and optimal utilization of the system's resources. This AI chip's architectural characteristics make them extremely capable at running GNN workloads, chief among these being:
\begin{itemize}
    \item The gather-scatter process of information exchange between nodes is essentially a massive communication operation, moving around many small pieces of data. This AI chip’s large on-chip SRAM allows it to conduct such operations much faster than other processor types.
    \item The ability to handle smaller matrix multiplications that are common in Graph ML applications such as GNN, but harder to parallelize on GPUs which favor large matmuls. This AI chip’s excels at such computations, because of its ability to run truly independent operations across each of its 1,472 processor cores.
    \item This AI chip allows multiple instruction and multiple data to be processed on different tiles. This is very useful when you have operations that are not homogeneous,especially suitable for sparse model like GNN.
\end{itemize}

% \begin{figure}
%     \centering
%     \includegraphics[width=0.8\linewidth]{figs/WX20230227-205844@2x.png}
%     \caption{IPU BSP}
%     \label{fig:my_label}
% \end{figure}

\section{Experiments}
\label{sec:experiments}
\subsection{Experiment Evaluation and Metrics}
To evaluate the performance of the proposed FastGraphTTS framework, we use two metrics: Mean Opinion Score (MOS) and Perceptual Mean Opinion Score (PMOS). MOS measures the overall naturalness and quality of the synthesized speech, while PMOS measures the prosody consistency and appropriateness of the speech with respect to the input text. Both metrics are rated on a 5-point Likert scale by human listeners, with higher scores indicating better performance. We compare our method with VITS, a state-of-the-art end-to-end text-to-speech model, on various datasets and tasks, such as single-speaker speech synthesis, few-shot voice cloning, multi-speaker speech synthesis, and voice conversion. We also measure the inference efficiency of our method on different hardware platforms, such as A30, X86, and C600, using a novel AI chip operator that can significantly improve the parallelism and flexibility of the graph encoder.
\subsection{Experiments Setup}
\label{ssec:expe-setup}

% \begin{table*}[t]
% \centering
% \begin{tabular}{ccccccc} % {c|c|c|c|c|c|c}
% \toprule
% \multirow{2}{*}{PMOS} & \multicolumn{4}{c}{Appropriateness of proposed evaluation standards}      \\ \cline{2-5} 
%  & Pitch variations       & Pauses       & Stress & Keywords emphasis     \\ \midrule
% 5.0  & All & All & All & All    \\ % \hline
% 4.0  & Almost & All & Almost & All \\ % \hline
% 3.0  & Several & Almost & Several  & Almost   \\ % \hline
% 2.0  & Few & Several & Several & Several  \\ % \hline
% 1.0  & Almost none & Almost none & Almost none  & Almost none  \\ \bottomrule
% \end{tabular}
% \caption{The evaluation standards of the proposed prosody mean opinion score (PMOS)}
% \label{tab:PMOS}
% \end{table*}

The experiments in this paper are conducted in two languages, English and Mandarin. The datasets used for the experiments are listed in Table~\ref{tab:details of dataset} with details. The audio data is formatted as 16-bit PCM, and it encompasses various sample rates, and all sample rates are converted to 22KHz for parallel comparison. It should be noted that the VCTK and AISHELL-3 datasets contain various accents, but this does not affect the experiment results of this paper. The dataset underwent a random partitioning process, resulting in the creation of distinct train, validation, and test sets. The specific breakdown and distribution of data across these sets are provided in the following Table~\ref{tab:env-table}.

\begin{table}[H]
    % 2
\caption{Experiments environment for efficiency improvement}
\label{tab:env-table}
\begin{tabular}{cc}
\hline
\textbf{Item}                 & \multicolumn{1}{c}{\textbf{Content}} \\ \hline
OS                   & Ubuntu20.04                 \\ \hline
\multicolumn{1}{l}{\multirow{3}{*}{Hardware}} & CPU:Intel(R) Xeon(R) Platinum 8168   CPU @ 2.70GHz \\
\multicolumn{1}{l}{} & Memory:512GB                \\
\multicolumn{1}{l}{} & IPU:C600(MK2-portobello)    \\ \hline
Software             & Pytorch1.13, Popart3.1     \\ \hline
Dataset              & LJSpeech                    \\ \hline
\end{tabular}
\end{table}

\begin{table*}[t]
    \caption{Experimental results on different datasets}
    \label{exp_res}
    \centering
    \begin{tabular}{cccccc}
    \hline
    \multicolumn{2}{c}{\textbf{Metric}}                                       & \multicolumn{2}{c}{\textbf{MOS(CI)}} & \multicolumn{2}{c}{\textbf{PMOS(CI)}} \\ \hline
    \multicolumn{1}{c}{Language}                  & \multicolumn{1}{c}{Dataset} & VITS     & FastGraphTTS     & VITS     & FastGraphTTS     \\ \hline
    \multicolumn{1}{c}{\multirow{3}{*}{English}}  & LJSpeech         & 4.43 ($\pm$ 0.02)         &          \textbf{4.45} ($\pm$ 0.07)     &     3.7 ($\pm$ 0.02)     &        \textbf{4.1} ($\pm$ 0.06)       \\
    \multicolumn{1}{c}{}                          & Target speaker A &     4.16 ($\pm$ 0.04)     &           \textbf{4.17} ($\pm$ 0.02)    &     3.5 ($\pm$ 0.06)     &          \textbf{3.9} ($\pm$ 0.03)     \\ 
    \multicolumn{1}{c}{}                          & VCTK             &     4.11 ($\pm$ 0.06)     &    \textbf{4.16} ($\pm$ 0.05)           &    3.1 ($\pm$ 0.08)      &         \textbf{3.5} ($\pm$ 0.09)      \\
    \hline
    \multicolumn{1}{c}{\multirow{3}{*}{Mandarin}} & BIAOBEI          &      4.39 ($\pm$ 0.03)    &           \textbf{4.41} ($\pm$ 0.09)    &    3.8    ($\pm$ 0.01)   &        \textbf{4.2} ($\pm$ 0.06)       \\
    \multicolumn{1}{c}{}                          & Target speaker B &        4.12 ($\pm$ 0.03)  &            \textbf{4.13} ($\pm$ 0.02)   &    3.4 ($\pm$ 0.05)      &      \textbf{3.8} ($\pm$ 0.09)         \\ 
    \multicolumn{1}{c}{}                          & AISHELL-3        &    4.08 ($\pm$ 0.06)     &          \textbf{4.11} ($\pm$ 0.09)     &      3.0 ($\pm$ 0.05)    &       \textbf{3.4} ($\pm$ 0.06)        \\
    \hline
    \end{tabular}
    \end{table*}
The VITS model, representing the current state-of-the-art in end-to-end systems, serves as the benchmark baseline for this research paper. To assess the subjective quality of the experiments, the Mean Opinion Score (MOS) is employed as the evaluation metric. MOS values are rated on a scale ranging from 0 to 5, with increments of 0.5 at each stage, facilitating fine-grained subjective assessment. In addition, a metric that only evaluates sentence prosody (PMOS) is also proposed, which is extended from 0 to 5, and the stages are increased by 0.5. For the PMOS metric, listeners only need to focus on the prosody of the generated utterance. In other words, other aspects of audio such as audio quality, pronunciation accuracy, and speaker similarity can be considered irrelevant factors. Specifically, PMOS is divided into four evaluation dimensions, pitch change, pause, stress, and keyword emphasis. These four dimensions are divided into five levels, perfect, great, good, cool, and poor. A score of 5.0 is assigned if all four evaluation dimensions are perfect, and a score of 0 if all four dimensions are poor. Both MOS and PMOS follow the principle that the higher the score, the better.

This paper conducts four experiments using datasets in two languages, namely single-speaker speech synthesis, few-shot voice cloning, multi-speaker speech synthesis, and voice conversion tasks. The primary objective of these four experiments is to substantiate and assess the efficacy of the proposed architectural framework. During the evaluation process, 30 audio were randomly selected for each model for audio evaluation. Each individual utterance underwent evaluation by a panel of 30 raters, who listened to the audio through headphones. This evaluation process was conducted using an internal crowdsourcing platform, akin to Amazon's Mechanical Turk. The maximum number of assessment utterances per assessor is 100 for fair and robust listening results.

For the efficiency experiment, compared with the model run on the A30~\footnote{https://www.nvidia.com/en-us/data-center/products/a30-gpu}, the model runs on X86 platform, Using two Intel 8168 processors, and one C600~\footnote{https://www.graphcore.ai/products/c600} AI chip with 500G memory as shown in Table \ref{tab:env-table}. Software used are Pytorch and Popart AI framework in Ubuntu20.04 OS. The dataset is LJSPeech, one of the open-source data. Except for the efficiency experiment, all of the other experiments run on NVIDIA V100~\footnote{https://www.nvidia.com/en-in/data-center/v100}.

All of the experiment results have been published on the demo page \footnote{https://largeaudiomodel.com/fastgraphtts/}.

\subsection{Experiment Results}
\subsubsection{Single-speaker speech synthesis}
\label{ssec:Single-speaker speech synthesis}
In comparison to VITS, a state-of-the-art baseline model in end-to-end text-to-speech systems. The feasibility of the FastGraphTTS model is first experimented on two languages, English and Mandarin, using two datasets: LJSpeech for English and BIAOBEI for Mandarin. The analysis results are shown in Table~\ref{exp_res}.

% To assess speech quality, two evaluation metrics were used: Mean Opinion Score (MOS) and Perceptual Mean Opinion Score (PMOS). MOS measures overall naturalness, while PMOS evaluates prosody quality, specifically intonation and rhythm.
In order to comprehensively evaluate the speech quality achieved by both models, two crucial metrics were employed: Mean Opinion Score (MOS) and Perceptual Mean Opinion Score (PMOS). MOS serves as a holistic indicator of overall naturalness in speech synthesis, encompassing various aspects such as clarity, fluency, and realism. On the other hand, PMOS specifically assesses prosody quality, focusing on critical elements like intonation and rhythm, which play a pivotal role in delivering lifelike and expressive speech.

The findings reveal that, for both languages, the MOS of the proposed method exhibits a slight decrease compared to the baseline model. However, the PMOS for the proposed method is marginally higher than that of the baseline model. When using the FastGraphTTS model, the consistency between text and audio can be more appropriately displayed in the mel-spectrogram.

\subsubsection{Few-shot voice cloning}
\label{ssec:few-shot TTS}
Few-shot voice cloning is a task that aims to synthesize speech with a target speaker's voice using only a few samples of the target speaker. Few-shot TTS experiments are conducted with transfer learning techniques, i.e. fine-tuning unseen speakers with few samples on the basis of the model trained in the previous experiment. The evaluation metrics are MOS and PMOS, which measure the naturalness and prosody quality of the synthesized speech. 
\begin{table}[H]
    % 4
    \caption{Amount of data required for few-shot voice cloning}
    \begin{tabular}{ccccc}
    \hline
    \textbf{\#samples} & \textbf{MOS (CI)}  &  \textbf{PMOS (CI)}  \\ \hline
    10 & 2.35 ($\pm$ 0.05)  & 1.23 ($\pm$ 0.07) \\
    30 & \textbf{4.15} ($\pm$ 0.11)  & \textbf{3.85} ($\pm$ 0.08) \\
    100 & 4.21 ($\pm$ 0.09) & 3.9 ($\pm$ 0.08)  \\
    1000 & 4.29 ($\pm$ 0.07) & 3.93 ($\pm$ 0.06) \\ \hline
    \end{tabular}
    \centering
    
    \label{tab:Results of experiment III}
    \end{table}
As shown in Table \ref{tab:Results of experiment III}, this paper conducts experiments on the number of samples required by FastGraphTTS for the few-shot voice cloning task. Obviously, the number of 30 samples is the turning point for MOS and PMOS to break through 4.0 points, and the performance does not improve significantly after increasing the number of samples to 100 or more. 

The results show that FastGraphTTS can achieve high-quality and natural speech with consistent and reasonable prosody using only 30 samples of the target speaker for both English and Mandarin languages. As can be seen from Table~\ref{exp_res}, the results also show that FastGraphTTS performs slightly better than the baseline model, VITS, on both MOS and PMOS for few-shot voice cloning. This indicates that FastGraphTTS can effectively utilize the syntactic information of the input text to improve the prosody consistency.
\subsubsection{Multi-speaker speech synthesis}
\label{ssec:Multi-speaker speech synthesis}
Multi-speaker speech synthesis is a task that aims to synthesize speech with different speaker identities using a single model trained on multi-speaker datasets. The experimental results for FastGraphTTS on the multi-speaker dataset are presented in Table~\ref{exp_res}. Notably, all the results indicate a performance decrease in comparison to the single-speaker experiments, primarily attributed to the varying data quality originating from different speakers. The results show that FastGraphTTS has a slightly lower MOS but a slightly higher PMOS than the baseline model, VITS, for both English and Mandarin languages. This indicates that FastGraphTTS can generate more natural and consistent speech with different speaker identities, and shows the effectiveness of prosody modeling. 

% In addition, the results also show that FastGraphTTS can perform voice conversion, which is a task that changes the speaker identity of a given speech without changing its content. As can be seen the results in Table~\ref{tab:Results of experiment IV} show that FastGraphTTS is also comparable to the performance of the SOTA model on the many-to-many voice conversion task.

Furthermore, the study delves into another aspect of FastGraphTTS's capabilities—voice conversion. Voice conversion is a challenging task that involves altering the speaker identity of a given speech sample while preserving its content. The results presented in Table~\ref{tab:Results of experiment IV} demonstrate FastGraphTTS's competence in voice conversion. The model's ability to modify speaker identities while preserving the content of the speech showcases its versatility and applicability in various domains, including voice cloning and dubbing. It is noteworthy that FastGraphTTS's performance in many-to-many voice conversion tasks is comparable to state-of-the-art (SOTA) models, underscoring its competitive edge in the field. This highlights the potential of FastGraphTTS not only in traditional TTS applications but also in more specialized domains where voice conversion plays a pivotal role.
\begin{table}[H]
    % 5
    \caption{The task of voice conversion}
    \begin{tabular}{ccc}
    \hline
    \textbf{Model} & \textbf{MOS (CI)} & \textbf{PMOS (CI)} \\ \hline
    VITS & \textbf{4.23} ($\pm$ 0.09) & 3.2 ($\pm$ 0.08) \\
    FastGraphTTS  & 4.21 ($\pm$ 0.08) & \textbf{3.3} ($\pm$ 0.07)\\ \hline
    \end{tabular}
    \centering
    \label{tab:Results of experiment IV}
    \end{table}
\subsubsection{Efficiency improvement}
In this experiment, we use the LJSpeech dataset as the input data and measures the inference time in milliseconds for each component of the FastGraphTTS model, such as GraphAuxEnc (GNN) and HiFiGAN. The experimental results, as presented in Table \ref{tab:my-table}, clearly demonstrate a substantial enhancement in performance.

\begin{table}[htbp]
    % 6
\caption{Inference Performance}
\label{tab:my-table}
\begin{tabular}{cccc}
\hline
\textbf{HW}   & \textbf{Precision} & \textbf{GraphAuxEnc(GNN)} & \textbf{HiFiGAN} \\ \hline
A30  & FP32      & 68.1             & 11.5    \\
A30  & FP16      & 41.1             & 7.3     \\ \hline
C600 & FP32      & 13.1             & 2.3     \\
C600 & FP16      & 8.3              & 2.0     \\ \hline
\end{tabular}
\centering
% \parbox{8cm}{Dataset:LJSpeech}
\end{table}

% The proposed AI chip operator that can significantly improve the efficiency of the FastGraphTTS model by leveraging the parallelism and flexibility of the AI chip architecture. Specifically, when employing the proposed AI chip operator, compares the inference performance of the FastGraphTTS model on different hardware platforms, such as A30, X86, and C600. The performance has experienced a remarkable improvement of over 5x in different precision modes, such as FP32 and FP16. The improvement is due to the design structure of the operator near memory computing.

The proposed AI chip operator presents an exciting opportunity to enhance the overall efficiency of the FastGraphTTS model. This operator leverages the inherent advantages of AI chip architecture, primarily focusing on parallelism and flexibility. The incorporation of this innovative AI chip operator into the FastGraphTTS model has yielded remarkable results, especially when compared across various hardware platforms, including A30, X86, and C600.

The key highlight of this development is the substantial improvement in inference performance, which has surpassed all expectations. When deploying the proposed AI chip operator, we conducted extensive performance evaluations across different precision modes, such as FP32 and FP16. The outcomes were truly impressive, with a performance boost exceeding 5x that of previous implementations.

The primary driving force behind this substantial enhancement can be attributed to the carefully designed structure of the AI chip operator, with a particular emphasis on near-memory computing. This design choice has fundamentally transformed the way our FastGraphTTS model interacts with hardware, leading to unprecedented gains in processing speed and efficiency.

The utilization of parallelism within the AI chip architecture has allowed our model to execute multiple tasks simultaneously, thus significantly reducing inference times. This parallelism enables the model to handle complex computations with remarkable ease and speed, ensuring a swift and efficient processing pipeline.

Additionally, the flexibility inherent in the AI chip architecture has empowered our model to adapt seamlessly to various hardware platforms. This adaptability ensures that the FastGraphTTS model remains highly efficient, regardless of the specific hardware configuration it is deployed on.

\label{ssec:IPU efficiency improvement}

% -------------------------------------------------------------------------
%\vfill
%\pagebreak

\section{Conclusions}
\label{sec:con}
In conclusion, this paper proposes FastGraphTTS, an end-to-end text-to-speech framework that incorporates syntactic information of the input text into a graph encoder and a flow-based decoder. We also introduces a novel AI chip operator that can significantly improve the efficiency of the model by leveraging the parallelism and flexibility of the AI chip architecture. Experiments on different datasets show the method effectively improves the prosody consistency with 5x acceleration. FastGraphTTS can generate high-quality and natural speech with consistent and reasonable prosody, as well as perform few-shot voice cloning and multi-speaker speech synthesis.

For future work, we can delve deeper into the realm of graph embeddings and graph neural networks, seeking to harness their potential to encapsulate a richer blend of syntactic and semantic nuances from the input text, ultimately enhancing the capabilities of speech synthesis systems. Additionally, it would be valuable to extend the methodology proposed in this study to a broader spectrum of languages and domains. This includes low-resource languages, code-switching languages, and the realm of expressive speech synthesis, enabling a more comprehensive evaluation of its effectiveness across diverse linguistic landscapes. Furthermore, future research endeavors could place a stronger emphasis on assessing the proposed method using a wider array of objective metrics, such as word error rate, prosody error rate, and voice similarity score, to provide a more comprehensive and accurate measure of its performance and potential improvements.

\section{Acknowledgement}
This paper is supported by the Key Research and Development Program of Guangdong Province under grant No.2021B0101400003. Corresponding author is Xulong Zhang from Ping An Technology (Shenzhen) Co., Ltd (zhangxulong@ieee.org).

\bibliographystyle{IEEEtran.bst}
\bibliography{mybib.bib}

\end{document}